# Data Hiding in Binary Image using Block Parity

Sipendra Sinha, Amol Gaikwad, Deepak Kumar, Snehal Darade, Rohit Singh

Mr. Pramod D. Ganjewar
*Computer Engg. Department, MIT Academy of Engineering,Alandi,Pune*
Pune University
`pdganjewar@comp.maepune.ac.in`

*Abstract*— Secret data hiding in binary images is more difficult than other formats since binary images require only one bit representation to indicate black and white. This study proposes a new method for data hiding in binary images using optimized bit position to replace a secret bit. This method manipulates blocks, which are sub-divided. The parity bit for a specified block decides whether to change or not, to embed a secret bit. By finding the best position to insert a secret bit for each divided block, the image quality of the resulting stego-image can be improved, while maintaining low computational complexity.
The experimental results show that the proposed method has an improvement with respect to a previous work.

## I. INTRODUCTION

The security of digital media becomes of major concern due to its emergence and wide spread use. The security of the transformation of hidden data can be achieved by two ways: encryption and steganography. A combination of the two techniques can be used to further increase the security of data. In encryption, the message is changed so that no data can be disclosed if received by an attacker.

In the proposed scheme, the binary (black and white) image is used for hiding the secret message. Hiding in a binary image is very difficult due to the fact that changing one bit in a binary image is easy to detect as it changes the color from black to white or the opposite. However, there are many techniques for a hiding a message in a binary image .This project presents a new scheme for embedding four bits in 5×5 block in a binary image by changing at most two bits with an efficient and invisible way without using a key.

## II. Related Work

The CPT-algorithm is a well-known algorithm applied on a binary cover image to hide data. This algorithm is designed to embed at most log(mn+1) bits of the secret message in a m*n block by changing a maximum of two bits. Before discussing the algorithm, the following notations are defined.
• K: Secret key used to embed and extract the message. It has a size of m×n, and contains 0 & 1.
• W: Weight matrix also used in embedding and extracting the message. Its size is m×n and its numbers range from 1 to $2^r -1$.
• r: Number of bits embedded in one block.
• F: Covered image used to hide data.

1) The Embedding Algorithm

Input: Cover image (F), secret key (K), weight matrix (W) and secret message.

Output: Stego-image.
Step1: Divide F into blocks (Fi), each of size m×n.
Step2: Determine the size of r where
r ≤ ⌊log(mn+1)⌋
Step3: Perform the following steps for each Fi until the whole secret message is embedded:
a) Find (Fi XOR K) XOR W
b) Find the sum (s) of the matrix obtained by the previous step.
c) Find s mod m where $m=2^r$
d) Get the sequence of bits from the secret message.
e) Compare the values obtained from c and d: if they are equal, no action is required, else, one or two bits in Fi should be changed to make the values equal.

2) Extraction Algorithm
Input: Stego-image, secret key (K) and weight matrix (W).

Output: Secret message.
Step1: Divide F into blocks (Fi), each of size m*n.
Step2: Determine the size of r.

Step3: Perform the following steps for each Fi until the whole secret message is extracted:
a) Find (Fi XOR K) XOR W
b) Find the sum (s) of the matrix obtained by the previous step.
c) Find s mod m where $m=2^r$. The result is r bits of the secret message. However, the CPT algorithm can embed 4 bits in a 5×5 block using a secret key and a weight matrix. The proposed algorithm takes a fixed block size 5×5 that can embed 4 bits by changing a maximum of 2 bits as in the CPT algorithm without using a secret key or a weight matrix as described in the following section.

### III. PROPOSED METHOD

The binary image selected for our project is in PBM format, where each pixel represent black or white. Binary data has the byte order of little endian, so we have to reserve bits of each byte, while reading it. If the width of the image is not a multiple of 8, then last byte containing remaining bits, are padded with the 0's at the end. ASCII representation of such file is 8 time larger in size.

For our project we are using portabe bitmap format or PBM format. Each file starts with a two-byte magic number (in ASCII) that explains the type of file it is. The data is bi-level so we are using the magic number P4.

Binary File Structure
- Magic Number
- Comments Section
  comments are optional and starts with '#'.
- Width and height are in next line and are separated with a space.
- Binary Data

Embedding Algorithm :
1. We do row-wise extraction 5X5 blocks from binary as shown below.

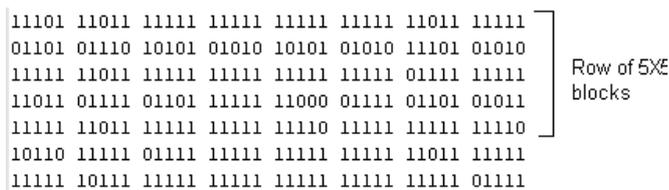

The whole image data is represented by a 4D array.

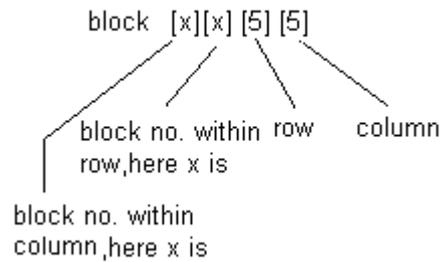

2. X-OR operations
For each block, except single-value black and white ones, proceed as follows:
1. For every row in the first four rows of the block, exclusive-or all the bits of that row to get r1r2r3r4.
2. For every column in the first four columns of the block, exclusive-or all the bits of that column to get c1c2c3c4.
3. Exclusive-or the results in 1 and 2 to get s1s2s3s4 where s1=r1 XOR c1, s2=r2 XOR c2, and so on.
4. Compare the result obtained from 3 with the four embedded bits b1b2b3b4. If there is no difference, no change of bits in F is needed, otherwise, consider the following cases:
- if the difference in one bit bi, the bit [F] i,5 or [F]5,i should be changed
- else if difference in two bits bi and bj, then the bit [F] i,j or [F]j,i should be changed.
- else if difference in three bits bi, bj and bk, then the bits
  (( [F] i,j or [F] j,i) and ( [F]k,5 or [F]5,k)) or
  (( [F] i,5 or [F] 5,i) and ( [F]k,j or [F]j,k)) or
  (( [F] 5,j or [F] j,5) and ( [F]k,i or [F]i,k))
  should be changed
- else (difference in four bits bi, bj, bk and bm) then the bits
  (( [F] i,j or [F] j,i) and ( [F]k,m or [F]m,k)) or
  (( [F] i,m or [F] m,i) and ( [F]k,j or [F]j,k)) or
  (( [F] m,j or [F] j,m) and ( [F]k,i or [F]i,k))
  should be changed

The selection of the bit depends on the number of adjacent bits with the same value. The bit that has the least number of adjacent bits is selected. This is because it has a minimum effect on the cover image when it is changed.

Extracting Algorithm :
The algorithm used for extracting is similar to that used for embedding. It performs the following steps to give the embedded data.

Step1: Divide the cover image into blocks (F) each of size 5×5.
Step2: For each block, except single-value black and white ones, proceed as follows:
1. For every row in the first four rows of the block, exclusive-or all the bits of that row to get r1r2r3r4.
2. For every column in the first four columns of the block, exclusive-or all the bits of that column to get c1c2c3c4.
3. Exclusive-or the results in 1 and 2 to get the embedded bits s1s2s3s4 where s1=r1 XOR c1,s2=r2 XOR c2, and so on.

Security of Data :
We can define which row/column to ignore for each 5*5 block. We can have a random pattern for this example.

    Block seq.    Ignored bit
     1st    ->    4th bit
     2nd    ->    5th bit
     3rd    ->    2nd bit

This list of pairs for all blocks should be shared to extract data. This is implemented by changing the 2 conditions of hiding data by using the value of key (1 to 5) instead of 5.

- if the difference in one bit bi, the bit [F] i,key or [F]5,key should be changed.
- else if difference in three bits bi, bj and bk, then the bits
  - (( [F] i,j or [F] j,i) and ( [F]k,key or [F]key,k)) or
  - (( [F] i,key or [F] key,i) and ( [F]k,j or [F]j,k)) or
  - (( [F] key,j or [F] j,key) and ( [F]k,i or [F]i,k)) should be changed

If the data to be hidden is much less than the total available blocks for hiding,then the modifications required to hide the data should be equally distributed to minimise the visual perception of the changes. This accomplished by creating equal space between blocks in which data is to be hidden such that,such blocks are equally spaced in the whole image.

## VI. Application
1) **Covert Communication:**
A spy in a foreign country wants to send messages abroad. He needs to use local communication channels in order to send the messages. He should assume that the communication channel is monitored. Sending encrypted messages would raise suspicion and could result in cutting the access to the communication infrastructure. It is therefore in his best interest to hide the presence of communication at all. This could be solved using a clever steganographic protocol.

2) **Digital Watermarking:**
The author of a digital image wants to "sign" the image so that no one else can attribute the authorship of the image to himself. The signature cannot be appended to the image file, nor can it be visibly imprinted on the image because such signatures can be easily removed or replaced. Cryptographic digital signatures cannot be applied because images are to be viewed by others and, therefore, will be distributed "in plain". Cryptographic digital signatures can be used for authentication of a communication channel but cannot protect an image posted on a web page.

3) **Fingerprinting:**
Movies are distributed to different people (as in pay-per-view distribution system). One wants to identify those that make illegal copies and sell them. Other scenario includes distributing sensitive information (images, videos) to several deputies and trying to trace down a traitor who leaks information to the enemy. One cannot use visible (audible) marking because such would look suspicious and could be easily removed. The marks must be perceptually invisible and must be present in every frame or image that

**4) Adding Caption To image:**
Movie dubbing in multiple languages, subtitles, tracking the use of the data (history file). For
example, one copy of a movie can be distributed with subtitles in several languages. The VCR, DVD player, TV set, or other video device can access and decode the additional text (subtitles) in real time from each frame, and display it on the TV screen. Although this could be arranged by appending information rather than invisibly embedding it, bandwidth requirements and necessary format changes may not allow us to do so.

**5) Fraud Detection:**
An imaging device, such as a digital camera, digital video-camera, or a scanner marks an image with a unique, robust, secure watermark before it is saved on a flash card, DAT tape, small mechanical hard drive or sent to output to another device such as computer, video capture board, etc. Embedding watermarks into digital images with the intent to detect the place and extent of image modifications will play an important role in detecting digital frauds, and it can be used to establish a chain of custody in the court of law. Digital images cannot be currently used in the court of law as proofs because of the ease of making digital forgeries and the impossibility to detect image manipulation. The advantage of using digital watermarks is clear: the watermarks are independent of the image format, do not increase the bandwidth (as opposed to adding a header), and cannot be removed to prevent the proof of forgery.

V. EXPERIMNTAL RESULTS

The proposed algorithm together with the CPT algorithm have been applied on different images for different data. We have 3 criteria for measuring the experimental results

- **Similarity:** Between the stego-image and the original image

- **Average:** It is computed for each pixel depending on its neighbors. Then the average of pixel average values is also computed to test the consistency between each pixel and its neighbors.

- **Standard Deviation:** Compute the average for each pixel depending on its neighbors, and then compare it with the original image.

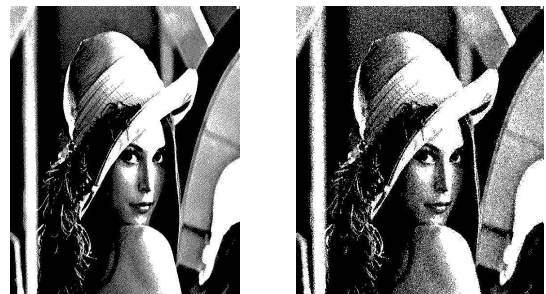

(a)Original Image     (b) Stego-Image

VI. Acknowledgement

We are very grateful to Prof. Pramod D. Ganjewar for helpful discussions on data hiding, including insights on visibility factor, noise and cryptanalysis techniques used to decipher image. We would also like to thank Atul Singh, Alok Shrivastava, and Ramesh chandra Gadri for helpful discussions on image distortion after hiding data, as well as their feedback on early versions of this work.

VII. CONCLUSION

The used algorithm does not need a secret key, it needs only an agreement between the embedding and extracting agents. Computational time is very less.Using of blocks reduces time of computation.The data hiding capacity is higher,as each 5X5 block can hide 4 bits of data.A standard image of dimensions 512 X 512 pixels can hide 5.08 KB of data.Extraction of data is faster as only 12 XOR operations are required to extract data from one block of data.